\documentclass[twocolumn,preprintnumbers,amsmath,aps]{revtex4-1}
\usepackage{graphicx}
\usepackage{dcolumn}
\usepackage{bm}
\usepackage{subfigure}
\usepackage{color}
\usepackage{microtype}
\begin{document}
\def\eq#1{(\ref{#1})}
\def\fig#1{figure\hspace{1mm}\ref{#1}}
\def\tab#1{table\hspace{1mm}\ref{#1}}
\title{Magnetic flux noise in superconducting qubits and the gap states continuum}

\author{Dominik Szcz{\c{e}}{\'s}niak$^{1, 2}$}\email{dszczesn@purdue.edu; d.szczesniak@ujd.edu.pl}
\author{Sabre Kais$^{1}$}

\affiliation{$^1$Department of Chemistry, Purdue University, 560 Oval Dr., 47907 West Lafayette, Indiana, United States of America}
\affiliation{$^2$ Department of Theoretical Physics, Faculty of Science and Technology, Jan D{\l}ugosz University in Cz{\c{e}}stochowa, 13/15 Armii Krajowej Ave., 42200 Cz{\c{e}}stochowa, Poland}
\date{\today}
\begin{abstract}

In the present study we investigate the selected local aspects of the metal-induced gap states (MIGSs) at the disordered metal-insulator interface, that were previously proposed to produce magnetic moments responsible for the magnetic flux noise in some of the superconducting qubit modalities. Our analysis attempts to supplement the available studies and provide new theoretical contribution toward their validation. In particular, we explicitly discuss the behavior of the MIGSs in the momentum space as a function of the local onsite energy deviation, that mimics random potential disorder at the interface. It is found, that when the difference between the characteristic electronic potentials in the insulator increases, the corresponding MIGSs become more localized. This effect is associated with the increasing degree of the potential disorder that was earlier observed to produce highly localized MIGSs in the superconducting qubits. At the same time, the presented findings show that the disorder-induced localization of the MIGSs can be related directly to the decay characteristics of these states as well as to the bulk electronic properties of the insulator. As a result, our study reinforces plausibility of the previous corresponding investigations on the origin of the flux noise, but also allows to draw future directions toward their better verification.

\end{abstract}
\maketitle
%

\section{Introduction}

The superconducting quantum interference devices (SQUIDs) are currently among the most promising platforms for the quantum information processing, that allow to build tunable qubits \cite{wendin, oliver}. However, although the superconducting approach satisfies all of the DiVincenzo's criteria, it still suffers from some important drawbacks \cite{gambetta1, oliver}. One of such limitations concerns the universal low-frequency $(1/f)$ magnetic flux noise that occurs in SQUIDs \cite{wellstood1, wellstood2, oliver} and greatly influences nearly all fundamental qubit modalities \cite{paladino, clarke}, such as the phase \cite{bialczak} and flux qubits \cite{yoshihara, kakuyanagi}. As a result, the magnetic flux noise visibly limits dephasing times of the mentioned qubit types \cite{bialczak, yoshihara, kakuyanagi}, hampers down their scalability \cite{paladino}, and reduces rate of coherently tunneling qubits in terms of the quantum annealers \cite{johnson, boixo}.

Over the years, various attempts to tackle the magnetic flux noise were proposed, that can be grouped into two main research strategies. First scenario concentrates on modifications of the superconducting qubit design, toward novel archetypes inherently insensitive to the magnetic noise \cite{vion, koch1, barends}. This approach is motivated by the elementary charge qubit, which does not suffer from the magnetic noise, yet it is susceptible to the charge fluctuations \cite{clarke}. In this context, new qubit types were introduced, such as the quantronium (via processes at noise unsusceptible bias points) \cite{vion}, transmon (based on shunt capacitive effect) \cite{koch1} or xmon (defined by the alternative capacitor geometries) \cite{barends}. However, despite being highly successful in some areas, this approach comes with still noticable trade-offs in terms of the intrinsic qubit anharmonicity, complicated control, and crowding in multi-qubit systems \cite{gambetta2, yan, sete, hutchings}. At the same time, it does not provide detailed understanding of the discussed magnetic noise, but rather attempts to bypass this issue. On the other hand, the second strategy aims at the in-depth elucidation of the magnetic flux noise in already existing superconducting qubit types, in order to suppress this negative phenomenon and improve performance of the related qubit solutions \cite{koch2, faoro1, faoro2, de, wu, choi}. In these terms, hitherto not fully explained origin of the magnetic noise in SQUIDs is still an intriguing and important aspect within the domain of superconducting qubits.

Interestingly, the magnitude of the magnetic flux noise is known to weakly depend on the SQUID area, as well as on the superconductor or substrate type \cite{bialczak}. According to that, local effects are argued to play important role in producing the discussed noise. In this regard, several theories were presented to explain the origin of the flux noise at the microscopic level. Specifically, Koch {\it et al.} proposed that the noise arises due to the electrons with random magnetic moments, that stochastically hop between defect centers at the surface of the superconductor \cite{koch2}. On the other hand, Faoro {\it et al.} attempted to explain this noise via dynamics of the spins that are strongly-coupled by the Ruderman-Kittel-Kasuya-Yosida (RKKY) interactions at the superconductor-insulator interface \cite{faoro1, faoro2}. In what follows, the RKKY interactions were also adopted in the spin-cluster model by De \cite{de}. Yet another proposal by Wu and Yu suggested that the noise emerges from the hyperfine interactions of the relaxing surface spins \cite{wu}. To this end, Choi {\it et al.} claimed increased role of the metal-induced gap states that become localized at the superconductor-insulator interface by the potential disorder \cite{choi}.

In general, the described above theories follow the most important experimental findings on the magnetic flux noise. Specifically, they render the universal and weak scalability of the noise with the overall size of the SQUID \cite{bialczak}. However, they also independently tackle other aspects of the discussed effect. In details, some of them emphasize the interactions between surface spins \cite{faoro1, faoro2, de, wu}, in agreement with the experimental observations provided in \cite{sendelbach2}. The other models \cite{choi}, concentrates on describing correctly the areal spin density from the susceptibility measurements \cite{sendelbach1}. Finally, the latest theoretical approaches attempt to include the role of an extrinsic effects coming from the surface adsorbents \cite{kumar, wang1}. As a result, they altogether suggest complex nature of the noise that may arise from the interplay of a various intrinsic and extrinsic effects.

In this context, there is a strong motivation to verify each of the theoretically postulated contributions to the origin of the flux noise. This will allow to account for the most important physical parameters that govern the noise, toward its further reduction. In the present study, of special interest is the role of the metal-induced gap states (MIGSs), as proposed by Choi {\it et al.} in \cite{choi}. The importance of the MIGSs steams from the fact that they are considered to inevitably appear whenever metal-insulator junction (MIJ) is created \cite{louie}. Thus, they should be also present at the Josephson junctions that build SQUIDs. Moreover, the MIGSs are likely to localize at the discussed interface, as they constitute direct continuum of the metal states that decay into the first few layers of the insulator \cite{tersoff}. According to Choi {\it et al.} \cite{choi}, the mentioned localization of the MIGSs notably contributes to the flux noise, giving rise to the paramagnetic local moments with the observed areal density \cite{sendelbach1}. As mentioned before, the localization should occur due to the presence of the potential disorder at the metal-insulator interface. Such interplay between the interfacial disorder and the localization of the MIGSs appears to be the central point of the approach presented by Choi {\it et al.} \cite{choi}. However, the universal and inherent nature of this relation has not been yet explicitly recognized, to further reinforce the importance of the MIGSs concept in terms of the magnetic flux noise.

In this study, we present new insight into the localization of the MIGSs, as caused by the variation of the site potential near the metal-insulator interface. Our analysis is based on the complex band structure method, that allows to directly investigate the behavior of the MIGSs in the momentum space within the energy gap. This way, the behavior of interest is determined from the insulator band structure, so that the localization effect is explicitly related to the intrinsic properties of the interface. Moreover, the calculations are employed at the local level to demonstrate importance of the relative energy difference, and its fluctuations, in correspondence to the fundamental features of the insulating layer. The presented here study can be viewed as a supplementary analysis to the investigations conducted by Choi {\it et al.} \cite{choi}, that considered behavior of the MIGSs within the scenario similar to the Cardona-Christiansen approximation \cite{cardona}.

\section{Theoretical Model}

To investigate the localization effect of the MIGSs near the MIJ, we concentrate our attention on the insulator region in a benchmark CsCl structure. The electronic properties of such structure are described in the framework of the mean-field Hubbard Hamiltonian given as:
\begin{eqnarray}
\label{eq01}
\nonumber
H &=& \sum_{i, \sigma} \varepsilon_{i} c^{\dagger}_{i, \sigma} c_{i, \sigma} - \sum_{\left< i,j \right>, \left<\left< i,j \right>\right>, \sigma } t_{i,j} \left(c^{\dagger}_{i, \sigma} c_{j, \sigma} + c^{\dagger}_{j, \sigma} c_{i, \sigma} \right) \\
&+& U \sum_{i, \sigma} \left(n_{i, \uparrow} \left< n_{i, \downarrow} \right> + n_{i, \downarrow} \left< n_{i, \uparrow} \right> \right),
\end{eqnarray}
where $\varepsilon_{i}$ is the electronic potential of the $s$-type orbital at the $i$-th site, equal to -4 and 2 eV for the undisturbed Cs and Cl atoms, respectively. In what follows, the $c_{i, \sigma}$ ($c^{\dagger}_{i, \sigma}$) operator creates (anihilates) electron at the $i$-th site with the spin $\sigma= \uparrow, \downarrow$, whereas $j$ denotes nearest-neighbors (NN) or next-nearest-neighbors (NNN) of $i$. Hence, the $t_{i,j}$ is the hopping energy set to -0.5 eV for both the NN and NNN cases. To this end, the $U$ parameter describes the on-site Coulomb repulsion, with $n_{i, \sigma}=c^{\dagger}_{i, \sigma} c_{i, \sigma}$ being the number operator. We take $U=3.25$ eV, to account for the well localized electrons, as suggested in \cite{choi}. Note, that above assumptions allow us to conduct our analysis on the same footing with the discussion provided by Choi {\it et al.} (see \cite{choi} for more details). Moreover, the assumed $U$ value is much greater than the superconducting pairing gap, hence our approach is expect to be valid also in the case when the attached metal is superconducting. Nonetheless, it is instructive to note here that we do not explicitly consider attached metal, by arguing the fact that MIGS should constitute intrinsic property of the insulator, according to the previous similar studies \cite{heine, tersoff, monch, reuter, szczesniak1}. This is to say, the interfacial behavior of MIGS is described by the insulator complex band structure (CBS), that can be generalized to the interface \cite{reuter}. Such approach is possible, since MIGS constitute direct analytical continuation of the propagating states in a metal \cite{dang, szczesniak1}. It also allows to trace, hitherto not considered, canonical aspects of the MIGS localization at the MIJ.

In particular, the MIGSs are calculated here by adopting the CBS method, that requires us to solve the following generalized eigenvalue problem:
\begin{eqnarray}
\label{eq02}
\nonumber
&&\left[\left[ \begin{array}{c c}
E\mathbf{I}-\mathbf{H}_{n} & \mathbf{H}_{n-1} \\
\mathbf{I} & 0
\end{array}\right] - \vartheta
\left[ \begin{array}{c c}
-\mathbf{H}^{\dagger}_{n-1} & 0 \\
0 & \mathbf{I}
\end{array}\right]
\right]\\
&&\times \left[ \begin{array}{c}
\psi_{n} \\
\psi_{n+1}
\end{array} \right]=0.
\end{eqnarray}
In Eq. (\ref{eq02}), the $\mathbf{H}_{n}$ and $\mathbf{H}_{n+1}$ matrices are the component Hamiltonian terms for the reference unit cell and its coupling to the neighboring cells, respectively. In this manner, the $\psi_{n}$ and $\psi_{n+1}$ describes the wave function coefficients accordingly associated with the $n$ and $n+1$ cell. These coefficients are chosen so that they satisfy the following phase relation $\psi_{n+1}=\vartheta\psi_{n}$, where $\vartheta$ denotes the generalized Bloch phase factor. The problem of Eq. (\ref{eq02}) is solved in the self-consistent iterative manner, to match the paramagnetic solutions of Eq. (\ref{eq01}). Note, that the paramagnetic behavior of the insulator is dictated by the experimental results of Sendelbach {\it et al.} \cite{sendelbach1}. For more technical details on the CBS method we refer to \cite{reuter, szczesniak2}.

\begin{figure*}[ht!]
\includegraphics[width=\textwidth]{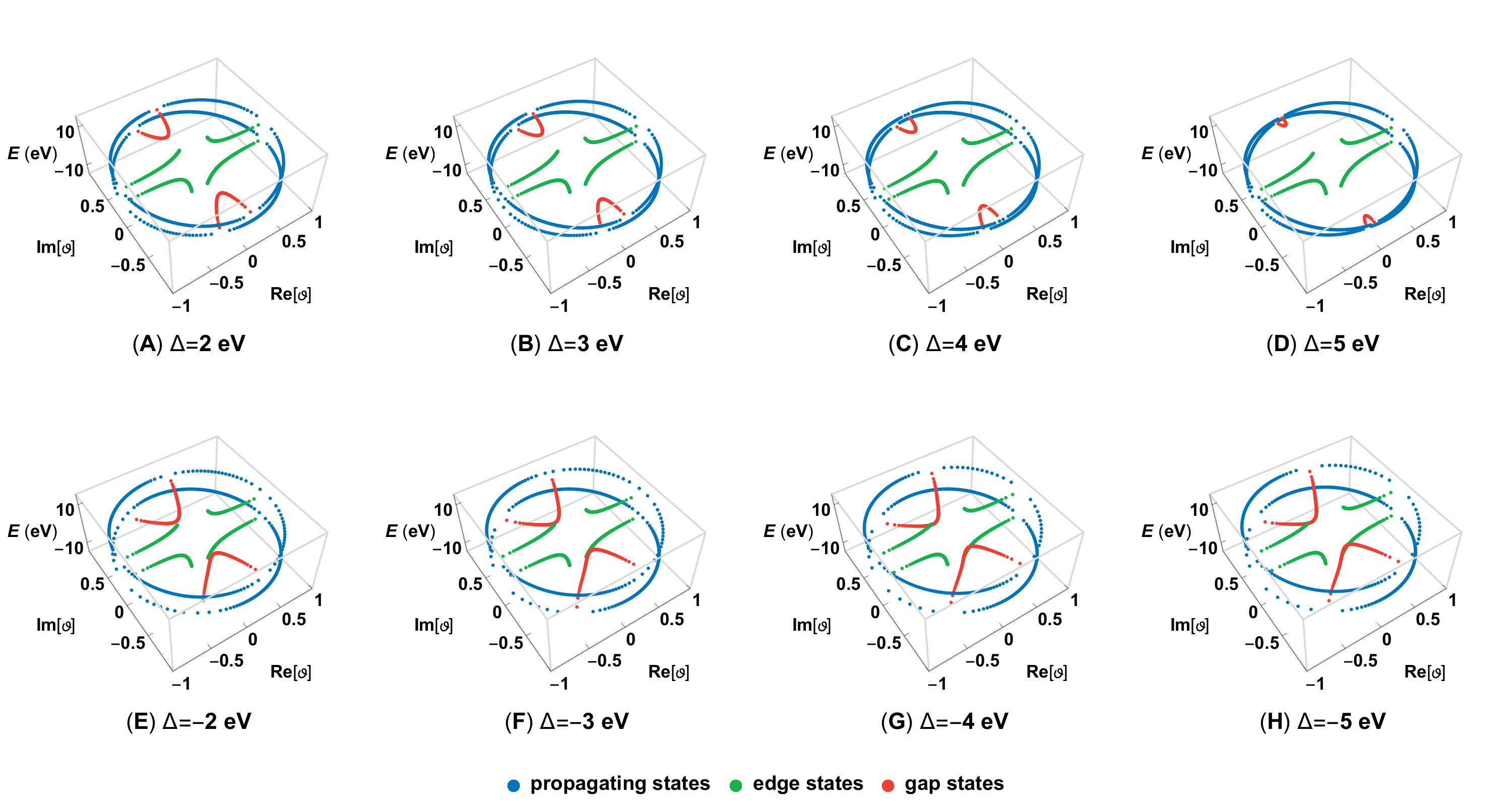}
\caption{The complex band structures of the CsCl insulator in the complex plane of generalized Bloch phase factor ($\vartheta$), for the selected values of the onsite energy deviation ($\Delta$). The first row of the sub-figures ((A)-(D)) depicts solutions for the positive values of $\Delta$, whereas the second row ((E)-(H)) presents results obtained for the negative $\Delta$ values. The propagating, edge and gap states are distinguished by the blue, green and red color, respectively.}
\label{fig01}
\end{figure*}

In general, the Eq. (\ref{eq02}) produces the pairs of $\vartheta$ and $1/\vartheta$ eigenvalues that are related to each other by the time-reversal symmetry. Herein, we consider the behavior of such solutions in the momentum space ($\mathbf{k}$). When $|\vartheta|=1$ the solutions correspond to the typical propagating states. On the other hand, when $|\vartheta|<1$ some of these eigenvalues can be interpreted as a MIGSs. Specifically, the MIGSs are the states with $|\vartheta|<1$ that appear in the energy gap and create characteristic complex band loops, joining the maximum of the valence band with the conduction band minimum. Of particular importance to the present discussion, is the point in energy space where the MIGSs exhibit the largest density. According to Choi {\it et al.}, these are the states most susceptible to the localization that should provide biggest contribution to the flux noise. By recalling the sum rule on the density of states \cite{appelbaum, claro}, the density of MIGSs must be derived from the contributions of the conduction and valence bands of the insulator. In this context, the highest density of MIGSs is expected at the point where the conduction and valence bands contributes equally to the MIGSs (the charge neutrality point). By following Allen \cite{allen1, allen2} and Tersoff \cite{tersoff}, this point can be located within the energy gap by using the following cell-averaged Green's function:
\begin{equation}
\label{eq03}
G(E, \mathbf{R})= \frac{1}{N} \sum_{n, \mathbf{k}} \frac{\vartheta}{E + i\eta - E_{n, \mathbf{k}}},
\end{equation}
where the parameter $N$ gives the number of the unit cells in the considered system, $n$ counts all the accessible eigenvalues in the momentum space ($E_{n, \mathbf{k}}$), and $\eta$ takes infinitesimal positive value to differentiate between the advanced and retarded Green's function. Moreover, the $\mathbf{R}$ is the lattice vector of the CsCl crystal structure. Since the function of Eq. (\ref{eq03}) varies its sign along with the bonding character of the considered eigenvalues, it yields the location of the charge neutrality point at the value of 0. The Im$[\mathbf{k}]$ value at this energy level is then interpreted as a characteristic decay rate ($\kappa$) of the MIGSs within the insulator energy gap; according to the wave function decay defined as $e^{- \kappa {\rm a}}$, with $a$ being the lattice constant. Hence, it provides the effective measure of how well given decaying states are localized near the MIJ.

In the context of the above, the present study attempts to analyze how the characteristic $\kappa$ value changes as a function of the potential disorder at the interface. However, to provide the most in-depth insight into the analyzed processes, their behavior is related here to the fundamental aspects of the potential disorder at the local level. In particular, when the random potential disorder is induced it varies the respective difference between the onsite energies ($\varepsilon_{i}$) at the two inequivalent sites (the Cs and Cl atoms) within each of the unit cells that build the insulator. This observation is of particular importance to the behavior of the MIGSs, since mentioned energy difference models region in the vicinity of the energy gap, as well as the gap itself. Therefore, it can be expected that this difference has particular influence on the behavior of the MIGSs. This fact is additionally reinforced by already mentioned observations made by Choi {\it et al.}, that suggest states located within the band gap or at its edges to be most susceptible to the localization. Note, however, that the presented approach is not limited to the local picture and can be extended further to analyze processes of interest in the framework of a large scale calculations {\it e.g.} within the approach derived from the Anderson localization model, as presented in \cite{choi}.

\section{Numerical results}

\begin{figure}[ht!]
\includegraphics[width=\columnwidth]{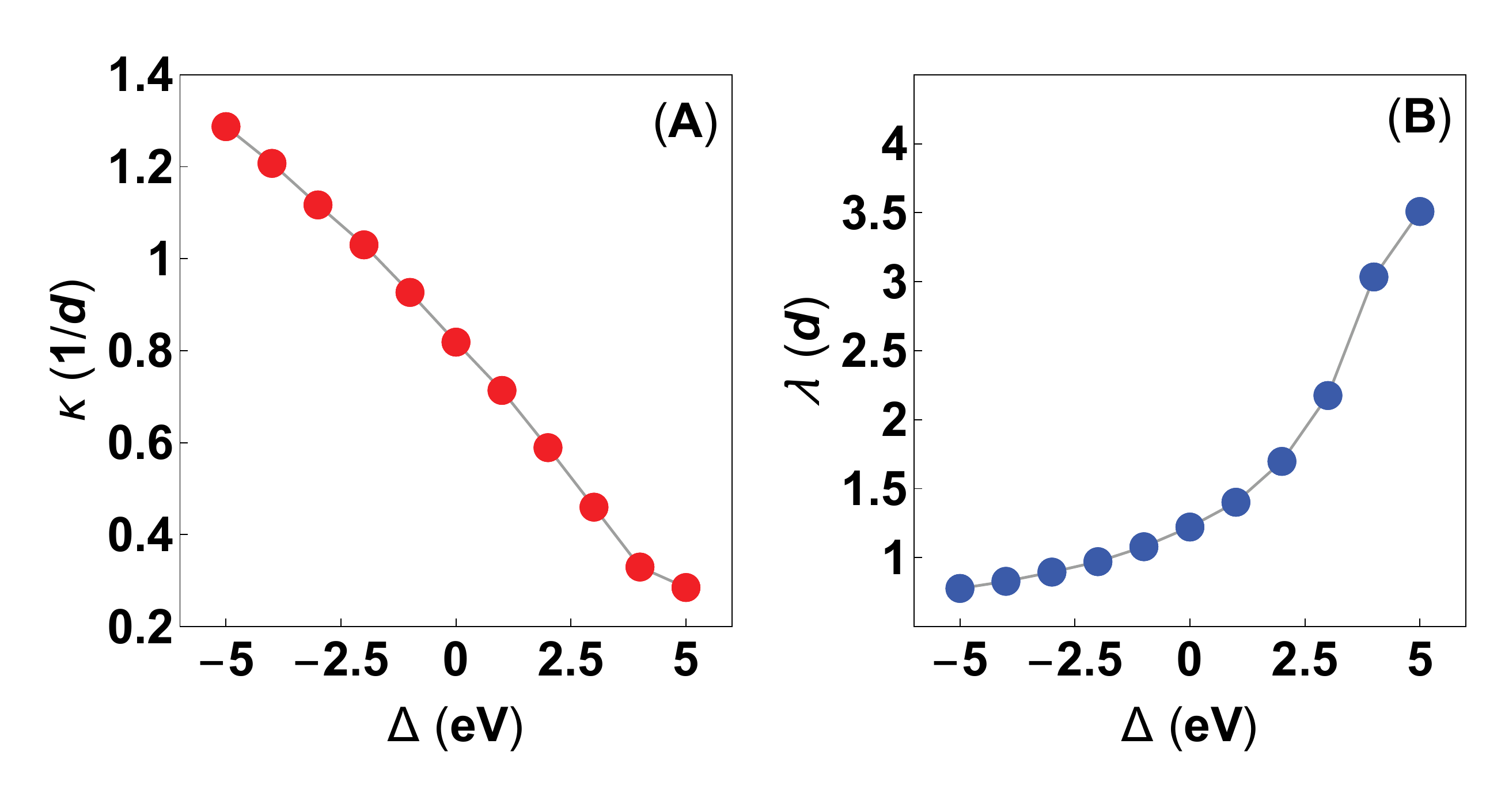}
\caption{(A) The characteristic decay rate ($\kappa$) and (B) corresponding decay length ($\lambda$) of the metal-induced gap states in the CsCl insulator structure, for the selected values of the onsite energy deviation ($\Delta \in [-5 {\rm eV}, 5 {\rm eV}]$). Note, that the $\kappa$ and $\lambda$ parameters are presented in the units that base on the primitive unit cell size ($d$) of the CsCl insulator structure.}
\label{fig02}
\end{figure}

In Fig. \ref{fig02} we depict the CBS solutions in the complex plane of $\vartheta$, that has been calculated based on Eq. (\ref{eq02}). The solutions are obtained for the selected deviations of the onsite energies within the insulator unit cell ($\Delta$). The first row of the sub-figures depicts solutions for the selected positive values of $\Delta$ (Fig. \ref{fig02} (A)-(D)), whereas the second row presented solutions obtained for the negative $\Delta$ values (Fig. \ref{fig02} (E)-(H)). As mentioned earlier, we argue here that the deviation $\Delta$ is the local manifestation of the potential disorder at the interface. Specifically, the energy deviation is defined as $\Delta \equiv \left| \varepsilon'_{{\rm Cs}} \right| - \left| \varepsilon''_{{\rm Cs}} \right| + \left| \varepsilon'_{{\rm Cl}} \right| - \left| \varepsilon''_{{\rm Cl}} \right|$, where $\varepsilon'_{{\rm Cs}}$ ($\varepsilon'_{{\rm Cl}}$) describes the unperturbed onsite energy at the Cs (Cl) atom, as given in Section II, and $\varepsilon''_{{\rm Cs}}$ ($\varepsilon''_{{\rm Cl}}$) denotes its perturbed counterpart. Hence, for $\Delta=0$ eV the solutions of Eq. (\ref{eq02}) describe the unperturbed band structure of the ideal CsCl insulator, and yield the characteristic electronic features of this material {\it e.g} the indirect band gap. Note finally that the reference zero energy in Fig. \ref{fig02} is set at the valence band maximum.

In general, the complex band structures in Fig. \ref{fig02} present set of three different types of states, that are characterized by the complicated functional behavior. Therein, the propagating states compose unit circles marked by the blue color ($|\vartheta|=1$). Note that these states are valid only deep into the bulk of the ideal periodic system. For the purpose of this study, however, we are only interested in the evanescent states that appear when the perfect periodicity of the system is broken {\it e.g.} at the surfaces, scattering incidents or the discussed here interfaces \cite{chang}. In what follows, we distinguish two major types of the evanescent states that can be observed within our theoretical model. First, we note the evanescent states that appear at the boundaries of the adopted Brillouin zone, that are depicted by the green color ($|\vartheta|<1 \cap {\rm Im[\vartheta]=0}$). These states exhibit exponentially decaying character inside the unit circles defined by the propagating states. Next, we observe the gap states of interest, that are given in the red color ($|\vartheta|<1 \cap {\rm Im[\vartheta]\neq0}$). Contrary to the boundary states they decay in the oscillatory manner and compose already mentioned complex band loops, that allows us to interpret them as a MIGSs. To elucidate the presented results even further we remind that each of the above solutions appears in pairs {\it i.e.} the pairs of the $\vartheta$ and $1/\vartheta$ eigenvalues. In details, the propagating and gap states are symmetric with respect to the origin of the Im$[\vartheta]$ axis, whereas the edge states exhibit symmetry around the Re$[\vartheta]$ axis. The component solutions of a given pair are interpreted as the right and left propagating (decaying) states, respectively. These solutions are related to each other by the time-reversal symmetry and holds the same information about given state, except of the propagation (decay) direction.

By inspecting the results given in Fig. \ref{fig02}, we can observe that the functional behavior of all states in the energy and $\vartheta$ space changes notably as the $\Delta$ parameter varies. Of particular importance to our discussion is the behavior of the gap states. One can notice that, as the $\Delta$ parameter decreases toward lowest negative values, the total magnitude of the complex band loops increases. In details, the size of the complex loops increases in both the $\vartheta$ complex plane as well as in the energy space. This behavior occurs due to the fact that the onsite energies at the Cs and Cl atoms determine the upper and lower energy limits of the complex band loops. Therefore, when their relative energy difference changes the size of the complex band loops changes accordingly. It can be then qualitatively argued that $\Delta$ strongly influences the MIGSs at the MIJ, including the pivotal position of the branch points associated with the complex band loops. However, to investigate how the $\Delta$ parameter is related to the localization strength of the MIGSs, their behavior must be analyzed in the momentum space.

In the context of the above, we adopt the cell-averaged Green's function method of Eq. (\ref{eq03}) to elucidate the relation between the MIGSs localization strength and the $\Delta$ parameter. Herein, we use the numerical techniques previously adopted in \cite{szczesniak1}. Specifically, by solving the Eq. (\ref{eq03}) for each of the presented cases, we search for the energy level ($E_{B}$) that corresponds to the location of the mentioned branch point at the given complex band loop. Next, we extract the corresponding value of the characteristic decay rate parameter, in each of the considered $\Delta$ cases, by using the following relation $\kappa= - \left| {\rm Re}[{\rm log}(\vartheta(E_{B}))/a]\right|$. The values of $\kappa$, determined in the described way, are presented in Fig. \ref{fig02} (A) as a function of $\Delta \in [-5 {\rm eV}, 5 {\rm eV}]$. Note that the decay rate parameter is expressed there in the units of the inverse unit cell size of the CsCl structure ($1/d$). In correspondence to the behavior of the complex band loops, the decay rate $\kappa$ decreases, almost linearly, along with the increase of the $\Delta$ parameter. In this manner, the highest values of $\kappa$ are obtained for the gap states with the biggest magnitude in the complex plane of $\vartheta$ (see Fig. \ref{fig01}). However, to derive the localization characteristics of the MIGSs from the results presented in Fig. \ref{fig02}, it is instructive to recall the fact that the wave function decay is given by $e^{- \kappa {\rm a}}$. With respect to this relation it is now obvious that states described by the highest values of $\kappa$ are the most localized ones. For an even better visualization of this aspect, one can define additional parameter known as the decay length. In particular, the decay length ($\lambda$) is simply given as the inverse of the decay rate parameter, namely $\lambda=1/\kappa$. In Fig. \ref{fig02} (B), the corresponding decay lengths of the discussed MIGSs are presented with respect to the $\Delta$ parameter. Note, that $\lambda$ is expressed now in the units of the unit cell size. In what follows, one can observe that the MIGSs with the largest $\kappa$ values are indeed the most localized ones, as they exhibit the shortest decay lengths.

The results presented in Fig. \ref{fig02} (A) and (B) allow us to draw several important observations regarding the localization of the MIGSs at the MIJ. Of particular importance is the fact that the provided results explicitly show direct relation between the MIGSs localization strength and the onsite energy deviation parameter ($\Delta$). By arguing the fact that $\Delta$ parameter is the local manifestation of the potential disorder at the MIJ, one can observe that certain interfacial variations of the potential may cause MIGSs to become more localized. In reference to the study of Choi {\it et al.} \cite{choi}, the energy distribution at the perturbed MIJ can be given as $P(E)=(1/\sqrt{2\pi} \delta) \times {\rm exp} \left[ -(E-E_{0})^{2}/2\delta^2 \right]$, where $E_{0}$ denotes the unperturbed onsite energy and $\delta$ is the standard deviation. In his work Choi {\it et al.} defines additionally the disorder degree as $R=2\delta/W$, with $W$ being the metal bandwidth. In the context of these relations, Choi {\it et al.} show that the localization of the MIGSs increases together with the increase of the $R$ parameter, that is proportional to the standard deviation. In our approach the onsite energy deviation ($\Delta$) practically sample, at the local level, the domain of the $\delta$ parameter. It can be then qualitatively argued that the highly localized MIGSs, observed by Choi {\it et al.}, correspond to the negative values of the $\Delta$ parameter in our analysis. On the other hand, the MIGSs described by the positive values of $\Delta$ are the states that decay over relatively large distances, and potentially do not contribute to the flux noise.

\section{Summary and conlusions}

In summary, in the present study we have employed the complex band structure method to elucidate local aspects of the MIGSs localization at the metal-insulator junctions, that occurs due to the potential disorder at the interface. The mentioned effect is of great importance, as it constitute the central point of the theory by Choi {\it et al.} \cite{choi} that attempts to explain the origin of the destructive flux noise in some of the superconducting qubit modalities. In this respect, our motivation was to supplement findings of the mentioned theory, but also to verify its plausibility from a new point of view.

In details, we have analyzed the behavior of the MIGSs in the momentum space, by associating them with the complex band structure solutions that appear within the energy gap of the benchmark CsCl insulator material. To mimic the random potential disorder near the interface at the local level, we have described this effect within the deviation of the onsite energies related to the two inequivalent sites within the CsCl unit cell. By varying the magnitude of the discussed deviation, it was found that the localization strength of the MIGSs notably changes. Specifically, we have shown that the MIGSs can indeed become more localized when the difference between onsite potentials at the Cs and Cl sites increases. As a result, it was suggested that these states may correspond to the highly localized MIGSs, observed by Choi {\it et al.} \cite{choi} under substantial degree of the interfacial disorder, that give rise to the local magnetic moments responsible for the flux noise in some superconducting qubits.

In what follows, in our analysis we have confirmed that the variations of the potential at the interface can cause stronger localization of the MIGSs, that appears as an effect directly related to the inherent electronic properties of the insulator material. At the same time, the obtained results validated, to some extent, the proposal of Choi {\it et al.}, saying that the MIGSs can play an important role in describing the magnetic flux noise in superconducting qubits. However, it should be noticed that further investigations are desirable, in order to combine our approach with the large scale modeling techniques, toward more comprehensive analysis of the discussed processes. In particular, it should be possible to adapt our theoretical techniques within the calculations that tackle realistic metal-insulator junctions with the random potential disorder. As least two directions in this respect can be listed: (i) the Anderson-derived scenario as considered by the Choi {\it et al.} \cite{choi} or (ii) the technique based on the renormalization group approach \cite{wang2}. Such investigations are expected to allow to relate the decay characteristics of the MIGSs, not only to the potential fluctuations, but also the experimentally observed areal density of the paramagnetic spins in SQUIDs. As a result, further verification of the flux noise origin in terms of MIGSs should be feasible. To this end, we also note that the described improvement of the theoretical techniques may allow to consider still open problems such as the role of the interactions between the paramagnetic spins in producing the flux noise.

\section{Acknowledgements}

D. Szcz{\c e}{\' s}niak acknowledges financial support of this work and the related research activities by the Polish National Agency for Academic Exchange (NAWA) under Bekker's programme (project no. PPN/BEK/2018/1/00433/U/00001). S. Kais would like to acknowledge funding by the U.S. Department of Energy (Office of Basic Energy Sciences) under award number DE-SC0019215.

\bibliographystyle{apsrev}
\bibliography{bibliography}

\end{document}